\documentclass[12pt]{article}
\usepackage[T1]{fontenc}
\usepackage{amsmath,amsfonts,amssymb}
\usepackage{theorem}
\newcommand{\Third}{{\textstyle{\frac{3}{2}}}}
\newcommand{\half}{{\textstyle{\frac{1}{2}}}}

\begin{document}
\newpage
\pagestyle{empty}
\vfill

\begin{center}

{\Large \textbf{\textsf{Sum Rules for Codon Usage Probabilities}}}

\vspace{10mm}

{\large L. Frappat$^{ac}$, A. Sciarrino$^{b}$, P. Sorba$^a$}

\vspace{10mm}

\emph{$^a$ Laboratoire d'Annecy-le-Vieux de Physique Th{\'e}orique LAPTH}

\emph{CNRS, UMR 5108, associ{\'e}e {\`a} l'Universit{\'e} de Savoie}

\emph{BP 110, F-74941 Annecy-le-Vieux Cedex, France}

\vspace{7mm}

\emph{$^b$ Dipartimento di Scienze Fisiche, Universit{\`a} di Napoli 
``Federico II''}

\emph{and I.N.F.N., Sezione di Napoli}

\emph{Complesso Universitario di Monte S. Angelo} 

\emph{Via Cintia, I-80126 Napoli, Italy}

\vspace{7mm}

\emph{$^c$ Member of Institut Universitaire de France}

\vspace{12mm}

\end{center}

\vspace{7mm}

\begin{abstract}
In the crystal basis model of the genetic code, it is deduced that the sum 
of usage probabilities of the codons with C and A in the third position for 
the quartets and/or sextets is independent of the biological species for 
vertebrates. A comparison with experimental data shows that the prediction 
is satisfied within about 5 \%.
\end{abstract}

\vfill
\vfill

\rightline{LAPTH-914/02}
\rightline{DSF-10/02}
\rightline{physics/0205013}
\rightline{May 2002}

\vspace*{3mm}
\hrule
\vspace*{3mm}
\noindent
\emph{E-mail:} \texttt{frappat@lapp.in2p3.fr}, 
\texttt{sciarrino@na.infn.it}, \texttt{sorba@lapp.in2p3.fr}

\newpage
\pagestyle{plain}
\setcounter{page}{1}
\baselineskip=16pt
 
The genetic code is degenerate. Degeneracy refers to the fact that almost 
all the amino acids are encoded by multiple codons. Degeneracy is found 
primarily in the third position of the codon, i.e. the nucleotide in the 
third position can change without changing the corresponding amino acid. 
Some codons are used much more frequently than others to encode a 
particular amino acid. The pattern of codon usage varies between species 
and even among tissues within a species [1--7].
The case of bacteriae has been widely studied \cite{bernardi85, osawa87}. 
To our knowledge, systematic studies for eukaryotes are rather fragmentary 
\cite{aota86, ikemura88, bernardi89}. Most of the analyses of the codon 
usage frequencies have adressed to analyse the relative abundance of a 
specified codon or in the comparison of the relative abundance in different 
biological species and/or genes. Little attention has been paid to 
correlations between codon usage frequency among different biological 
species. Indeed a correlation between suitable ratios of codon usage 
frequencies has been remarked in \cite{FSS1} for biological species 
belonging to the vertebrate class and in \cite{CFSS} for biological 
organisms including plants. Such correlations fit well in a mathematical 
model of the genetic code, called crystal basis model, proposed by the 
authors in \cite{FSS}. Moreover in \cite{FSS2} it has also been observed 
that the ratio of the previously defined quantities exhibits for 
vertebrates an almost universal behaviour, i.e. independent on the 
biological species and on the nature of the amino-acid, for the subset of 
the amino-acids encoded by quartets or sextets. These remarks suggest the 
possible existence of a general bias in the codon usage frequency of a 
specific codon belonging to a quartet or quartet sub-part of a sextet, i.e. 
the four codons differing for the last nucleotide. The aim of this paper is 
to investigate this aspect and to predict a general law which should be 
satisfied by all the biological species belonging to vertebrates.

\medskip

Let us define the usage probability for the codon $XZN$ ($X, Z, N \in \{A, 
C, G, U\}$) as
\begin{equation}
\label{eq:1}
P(XZN) = \lim_{n_{tot} \to \infty} \;\;\; \frac{n_{XZN}}{n_{tot}}
\end{equation}
where $n_{XZN}$ is the number of times the codons $XZN$ has been used in 
the biosynthesis process of the corresponding amino-acid and $n_{tot}$ is 
the total number of codons used to synthetise this amino-acid. It follows 
that our analysis and predictions hold for biological species with enough 
large statistics of codons. In our model each codon $XZN$ is described by a 
state belonging to an irreducible representation (irrep.), denoted 
$(J_{H},J_{V})^\xi$, of the algebra $U_{q}\big( sl(2)_{H} \oplus sl(2)_{V} 
\big)$ in the limit $q \to 0$ (so-called crystal basis); $J_{H}$, $J_{V}$ 
take (half-)integer values and the upper label $\xi$ removes the degeneracy 
when the same couple of values of $J_{H}$, $J_{V}$ appears several times. 
As can be seen in Table \ref{tablerep} there are for example four 
representations $(\half,\half)^\xi$, with $\xi=1,2,3,4$. Finally within the 
given representation $(J_{H},J_{V})^\xi$ two more quantum numbers 
$J_{H,3}$, $J_{V,3}$ are necessary to specify a particular state: see Table 
\ref{tablerep}, which is reported here to make the paper self-consistent.

\medskip

So it is natural, in the crystal basis model, to write the usage 
probability as a function of the biological species (b.s.), of the 
particular amino-acid and of the labels $J_{H}$, $J_{V}$, $J_{H,3}$, 
$J_{V,3}$ describing the state $XZN$. Here we suppose that the dependence 
of the amino-acid is completely determined by the set of labels $J's$ and 
so we write
\begin{equation}
\label{eq:2}
P(XZN) = P(b.s.; J_{H}, J_{V}, J_{H,3}, J_{V,3}) 
\end{equation}
Let us now make the hypothesis that we can write the r.h.s. of eq.  
(\ref{eq:2}) as the sum of two contributions: a universal function $\rho$ 
independent on the biological species and a b.s. depending function 
$f_{bs}$, i.e.
\begin{equation}
\label{eq:3}
P(XZN) = \rho^{XZ}(J_{H}, J_{V}, J_{H,3}, J_{V,3}) \; + \; 
f_{bs}^{XZ}(J_{H}, J_{V}, J_{H,3}, J_{V,3})
\end{equation}
As it is suggested by an analysis of the available data \cite{CFSS}, the 
contribution of $f_{bs}$ is not negligible but could be smaller than the 
one due to $\rho$. We will also assume
\begin{equation}
\label{eq:4}
f_{bs}^{XZ}(J_{H}, J_{V}, J_{H,3}, J_{V,3}) \approx F_{bs}^{XZ}(J_{H};J_{H,3}) 
\, + \, G_{bs}^{XZ}(J_{V};J_{V,3})
\end{equation}
Now, let us analyse in the light of the above considerations the usage 
probability for the quartets Ala, Gly, Pro, Thr and Val and for the quartet 
sub-part of the sextets Arg (i.e. the codons of the form CGN), Leu (i.e. 
CUN) and Ser (i.e. UCN). \\
For Thr, Pro, Ala and Ser we can write, using Table \ref{tablerep} and
eqs. (\ref{eq:2})-(\ref{eq:4}), with $N = A,C,G,U$,
\begin{equation}
\label{eq:5}
P(NCC) + P(NCA) = \rho_{C+A}^{NC} + F_{bs}^{NC}(\Third;x) + 
G_{bs}^{NC}(\Third;y) + F_{bs}^{NC}(\half;x') + G_{bs}^{NC}(\half;y')
\end{equation} 
where we have denoted by $\rho_{C+A}^{NC}$ the sum of the contribution of 
the universal function (i.e. not depending on the biological species) 
$\rho$ relative to $NCC$ and $NCA$, while the labels $x,y,x',y'$ depend on 
the nature of the first two nucleotides $NC$, see Table \ref{tablerep}. For 
the same amino acid we can also write
\begin{equation}
\label{eq:6}
P(NCG) + P(NCU) = \rho_{G+U}^{NC} + F_{bs}^{NC}(\Third;x) + 
G_{bs}^{NC}(\Third;y) + F_{bs}^{NC}(\half;x') + G_{bs}^{NC}(\half;y')
\end{equation}
Using the results of Table \ref{tablerep}, we can remark that the 
difference between eq.  (\ref{eq:5}) and eq.  (\ref{eq:6}) is a quantity 
independent of the biological species, 
\begin{equation}
\label{eq:7}
P(NCC) + P(NCA) - P(NCG) - P(NCU) \;=\; \rho_{C+A}^{NC} - \rho_{G+U}^{NC} 
\;=\; \mbox{Const.}
\end{equation}
In the same way, considering the cases of Leu, Val, Arg and Gly, we obtain 
with $W = C,G$
\begin{eqnarray}
\label{eq:8} 
P(WUC) + P(WUA) - P(WUG) - P(WUU) &=& \rho_{C+A}^{WU} - \rho_{G+U}^{WU} 
\;=\; \mbox{Const.} \\
\label{eq:9} 
P(CGC) + P(CGA) - P(CGG) - P(CGU) &=& \rho_{C+A}^{CG} - \rho_{G+U}^{CG} 
\;=\; \mbox{Const.} \\
\label{eq:10}
P(GGC) + P(GGA) - P(GGG) - P(GGU) &=& \rho_{C+A}^{GG} - \rho_{G+U}^{GG} 
\;=\; \mbox{Const.}
\end{eqnarray}
Since the probabilities for one quadruplet are normalised to one, from eqs.  
(\ref{eq:6})-(\ref{eq:10}) we deduce that for all the eight amino acids the 
sum of probabilities of codon usage for codons with last A and C (or U and 
G) nucleotide is independent of the biological species, i.e.
\begin{eqnarray}
\label{eq:11} 
P(XZC) + P(XZA) = \mbox{Const.} \qquad (XZ = NC, CU, GU, CG, GG) 
\end{eqnarray}
In order to check our proposed sum rules eq. (\ref{eq:11}) on experimental 
data, we have considered the data for species from the GenBank (release 
127.0 of Dec. 2001) for vertebrates with codon statistics larger than 
60.000, see Table \ref{tabledata1}. In Table \ref{tablePCA} we report the 
experimental data for the selected 21 vertebrates. The comparison shows 
that our predictions are verified within 3-6 \%, which is an amazing 
result. Moreover if one considers only the species with highest statistics 
(No. 1, 2, 3 in Table \ref{tabledata1}) the prediction is verified within 
about 3 \%.

In Table \ref{tablestatXZN} we report the mean value and the standard 
deviation of the probability of usage of the codons $XZN$ $(XZ = NC, CU, 
GU, CG, GG)$ computed over all biological species given in Table 
\ref{tabledata1}. It can be remarked that this probability shows a rather 
large spread which is surprisingly reduced when one makes the sum (compare 
Tables \ref{tablestatXZN} and \ref{tablestat}).

Moreover, if we assume that for sextets the functions $F$ and $G$ depend 
really on the nature of the encoded amino acid rather than on the 
dinucleotide, from the analysis of the content in the irreps of Table 
\ref{tablerep}, we derive in a completely analogous way as above that for 
the amino acid Ser the sum $P'_{C+A}(S) = P(UCA) + P(AGC)$ is independent 
of the biological species (note that we have normalised the probabilites by 
$P(UCA) + P(AGC) + P(UCG) + P(AGU) = 1$). The experimental data are in good 
agreement with this a priori surprising result, see last column 
$P'_{C+A}(S)$ of Table \ref{tablePCA}.

In Table \ref{tablestat} we have reported the statistical estimators (mean 
value, standard deviation and their ratio) for the probabilities given in 
eq. (\ref{eq:11}). It should be remarked that essentially two species: 
\textit{Danio rerio} (zebrafish) and \textit{Pan troglodytes} (chimpanze), 
differ sensibly from the average value for most amino acids. For the 
chimpanze, it may be a statistical fluctuation due to the relative low 
number of codons.

In order to evaluate how our results are statistically more significant 
than generic results, we have computed the sum of the probabilities of the 
codon usage $P(XZC) + P(XZU)$ and $P(XZC) + P(XZG)$ where $XZ$ denote the 
dinucleotides of the eight quartets/sextets. To save space we do not report 
here the corresponding tables, but only the values of the statistical 
estimators, see Table \ref{tablestat}. We see that for $P(XZC) + P(XZU)$ 
the standard deviation is a bit larger than the one for $P(XZC) + P(XZA)$ 
and even larger for $P(XZC) + P(XZG)$. This result fits also in our model 
as in the first case the probability differs from species independent 
factors, essentially for the presence of factors $G_{bs}(J_{V};J_{V,3})$, 
while in the last case, also for the presence of factors 
$F_{bs}(J_{V};J_{V,3})$.

\medskip

Now let us make two important remarks. \\
-- If we write for $\rho(J_{H}, J_{V}, J_{H,3}, J_{V,3})$, or equivalently 
$\rho_{C+A}^{NC}$, an expression of the type (\ref{eq:4}), i.e. separating 
the $H$ from the $V$ dependence, it follows that the r.h.s. of eqs. 
(\ref{eq:5}) and (\ref{eq:6}) are equal, and consequently the probabilities 
$P(NCC) + P(NCA)$ and $P(NCU) + P(NCG)$ should be equal, which is not 
experimentally verified. This means that the coupling term between the $H$ 
and the $V$ is not negligible for the $\rho(J_{H}, J_{V}, J_{H,3}, 
J_{V,3})$ function. \\
-- Summing equations (\ref{eq:5}) and (\ref{eq:6}) we deduce that the 
expression $F_{bs}^{NC}(\Third;x) + G_{bs}^{NC}(\Third;y) + 
F_{bs}^{NC}(\half;x') + G_{bs}^{NC}(\half;y')$ is actually not depending on 
the biological species. {From} eqs. (\ref{eq:5}) and (\ref{eq:6}) for 
different values of $N$ and for analogous equations for the other four 
quartets, we can derive relations between sums of $F_{bs}^{NC}$ and/or 
$G_{bs}^{NC}$ functions which are independent of the biological species.

\medskip

Let us emphasize our claim: we have remarked that the sum of the usage 
probabilities of two suitably choosen codons is, within a few percent, a 
constant independent on the biological species for vertebrates, which well 
fits in the framework of the crystal basis model. Of course one can restate 
the above results stating the sum of the probability of codon usage $XZC + 
XZA$ is not depending on the nature of the biological specie, without any 
reference to crystal basis model. However a deeper analysis of Table 
\ref{tablePCA} shows that $P_{C+A}$ for Pro, Thr, Ala, Ser and Gly is of 
the order of 0.62, for Leu and Val of the order 0.35 while for Arg is of 
order 0.52. In the crystal basis model the roots, i.e. the dinucleotide 
formed by the first two nucleotides of the first 5 amino acids belong to 
the same irrep. (1,1), the roots of Leu and Val belong to the irrep. (0,1), 
while the root of Arg belongs to the irrep. (1,0). This is an interesting 
result, especially for Pro whose molecule has a different structure than 
the others amino acids (Pro has an imino group instead of an amino group).

\medskip

It is natural to wonder what happens for other biological species. The 
green plants exhibits roughly the same pattern, but probably a more 
reliable analysis has to be performed considering a splitting into 
families. For invertebrates, the large number of existing biological 
species and the lack of data with sufficient diversity prevents from 
applying a similar analysis. The case of bacteriae is rather interesting. 
Eubacteriae seem to avoid this pattern of correlations. This may be the 
influence of selection effects which may be stronger or effective in 
shorter times in less complicated species. For bacteriae the G+C content 
varies in a wide range from 25 \% to 75 \%. Hence one can argue that 
biological species with large difference in the G+C content exhibit large 
difference in the correlation pattern discussed in this paper. However, 
using the Genbank data, one finds for eubacteriae no correlation between 
the G+C content and the value of the probabilities eq. (\ref{eq:11}).

\bigskip

\textbf{Acknowledgments:} Partially supported by M.U.R.S.T. (Italy) and 
M.A.E. (France) in the framework of french-italian collaboration Galileo 
2000 and for one of us (A.S.) by the M.U.R.S.T. through National Research 
Project \textsl{SINTESI 2000}.

\begin{table}[htbp]
\caption{The eukaryotic or standard code code. The upper label denotes 
different irreducible representations. }
\label{tablerep}
\footnotesize
\begin{center}
\begin{tabular}{|cc|cc|rr|cc|cc|rr|}
\hline
codon & amino acid & $J_{H}$ & $J_{V}$ & $J_{3,H}$ & $J_{3,V}$& codon & 
amino acid & $J_{H}$ & $J_{V}$ & $J_{H,3}$ & $J_{V,3}$ \\
\hline
CCC & Pro P & $3/2$ & $3/2$ & $3/2$ & $3/2$ & UCC & Ser S & $3/2$ & $3/2$ & 
$1/2$ & $3/2$ \\
CCU & Pro P & $(1/2$ & $3/2)^1$ & $1/2$ & $3/2$ & UCU & Ser S & $(1/2$ & 
$3/2)^1$ & $-1/2$ & $3/2$ \\
CCG & Pro P & $(3/2$ & $1/2)^1$ & $3/2$ & $1/2$ & UCG & Ser S & $(3/2$ & 
$1/2)^1$ & $1/2$ & $1/2$ \\
CCA & Pro P & $(1/2$ & $1/2)^1$ & $1/2$ & $1/2$ & UCA & Ser S & $(1/2$ & 
$1/2)^1$ & $-1/2$ & $1/2$ \\[1mm] \hline
CUC & Leu L & $(1/2$ & $3/2)^2$ & $1/2$ & $3/2$ & UUC & Phe F & $3/2$ & 
$3/2$ & $-1/2$ & $3/2$ \\
CUU & Leu L & $(1/2$ & $3/2)^2$ & $-1/2$ & $3/2$ & UUU & Phe F & $3/2$ & 
$3/2$ & $-3/2$ & $3/2$ \\
CUG & Leu L & $(1/2$ & $1/2)^3$ & $1/2$ & $1/2$ & UUG & Leu L & $(3/2$ & 
$1/2)^1$ & $-1/2$ & $1/2$ \\
CUA & Leu L & $(1/2$ & $1/2)^3$ & $-1/2$ & $1/2$ & UUA & Leu L & $(3/2$ & 
$1/2)^1$ & $-3/2$ & $1/2$ \\[1mm] \hline
CGC & Arg R & $(3/2$ & $1/2)^2$ & $3/2$ & $1/2$ & UGC & Cys C & $(3/2$ & 
$1/2)^2$ & $1/2$ & $1/2$ \\
CGU & Arg R & $(1/2$ & $1/2)^2$ & $1/2$ & $1/2$ & UGU & Cys C & $(1/2$ & 
$1/2)^2$ & $-1/2$ & $1/2$ \\
CGG & Arg R & $(3/2$ & $1/2)^2$ & $3/2$ & $-1/2$ & UGG & Trp W & $(3/2$ & 
$1/2)^2$ & $1/2$ & $-1/2$ \\
CGA & Arg R & $(1/2$ & $1/2)^2$ & $1/2$ & $-1/2$ & UGA & Ter & $(1/2$ & 
$1/2)^2$ & $-1/2$ & $-1/2$ \\[1mm] \hline
CAC & His H & $(1/2$ & $1/2)^4$ & $1/2$ & $1/2$ & UAC & Tyr Y & $(3/2$ & 
$1/2)^2$ & $-1/2$ & $1/2$ \\
CAU & His H & $(1/2$ & $1/2)^4$ & $-1/2$ & $1/2$ & UAU & Tyr Y & $(3/2$ & 
$1/2)^2$ & $-3/2$ & $1/2$ \\
CAG & Gln Q & $(1/2$ & $1/2)^4$ & $1/2$ & $-1/2$ & UAG & Ter & $(3/2$ & 
$1/2)^2$ & $-1/2$ & $-1/2$ \\
CAA & Gln Q & $(1/2$ & $1/2)^4$ & $-1/2$ & $-1/2$ & UAA & Ter & $(3/2$ & 
$1/2)^2$ & $-3/2$ & $-1/2$ \\[1mm] \hline
GCC & Ala A & $3/2$ & $3/2$ & $3/2$ & $1/2$ & ACC & Thr T & $3/2$ & $3/2$ & 
$1/2$ & $1/2$ \\
GCU & Ala A & $(1/2$ & $3/2)^1$ & $1/2$ & $1/2$ & ACU & Thr T & $(1/2$ & 
$3/2)^1$ & $-1/2$ & $1/2$ \\
GCG & Ala A & $(3/2$ & $1/2)^1$ & $3/2$ & $-1/2$ & ACG & Thr T & $(3/2$ & 
$1/2)^1$ & $1/2$ & $-1/2$ \\
GCA & Ala A & $(1/2$ & $1/2)^1$ & $1/2$ & $-1/2$ & ACA & Thr T & $(1/2$ & 
$1/2)^1$ & $-1/2$ & $-1/2$ \\[1mm] \hline
GUC & Val V & $(1/2$ & $3/2)^2$ & $1/2$ & $1/2$ & AUC & Ile I & $3/2$ & 
$3/2$ & $-1/2$ & $1/2$ \\
GUU & Val V & $(1/2$ & $3/2)^2$ & $-1/2$ & $1/2$ & AUU & Ile I & $3/2$ & 
$3/2$ & $-3/2$ & $1/2$ \\
GUG & Val V & $(1/2$ & $1/2)^3$ & $1/2$ & $-1/2$ & AUG & Met M & $(3/2$ & 
$1/2)^1$ & $-1/2$ & $-1/2$ \\
GUA & Val V & $(1/2$ & $1/2)^3$ & $-1/2$ & $-1/2$ & AUA & Ile I & $(3/2$ & 
$1/2)^1$ & $-3/2$ & $-1/2$ \\[1mm] \hline
GGC & Gly G & $3/2$ & $3/2$ & $3/2$ & $-1/2$ & AGC & Ser S & $3/2$ & $3/2$ 
& $1/2$ & $-1/2$ \\
GGU & Gly G & $(1/2$ & $3/2)^1$ & $1/2$ & $-1/2$ & AGU & Ser S & $(1/2$ & 
$3/2)^1$ & $-1/2$ & $-1/2$ \\
GGG & Gly G & $3/2$ & $3/2$ & $3/2$ & $-3/2$ & AGG & Arg R & $3/2$ & $3/2$ 
& $1/2$ & $-3/2$ \\
GGA & Gly G & $(1/2$ & $3/2)^1$ & $1/2$ & $-3/2$ & AGA & Arg R & $(1/2$ & 
$3/2)^1$ & $-1/2$ & $-3/2$ \\[1mm] \hline
GAC & Asp D & $(1/2$ & $3/2)^2$ & $1/2$ & $-1/2$ & AAC & Asn N & $3/2$ & 
$3/2$ & $-1/2$ & $-1/2$ \\
GAU & Asp D & $(1/2$ & $3/2)^2$ & $-1/2$ & $-1/2$ & AAU & Asn N & $3/2$ & 
$3/2$ & $-3/2$ & $-1/2$ \\
GAG & Glu E & $(1/2$ & $3/2)^2$ & $1/2$ & $-3/2$ & AAG & Lys K & $3/2$ & 
$3/2$ & $-1/2$ & $-3/2$ \\
GAA & Glu E & $(1/2$ & $3/2)^2$ & $-1/2$ & $-3/2$ & AAA & Lys K & $3/2$ & 
$3/2$ & $-3/2$ & $-3/2$ \\[1mm] \hline
\end{tabular}
\end{center}
\end{table}

\clearpage

\begin{table}[t]
\caption{Data for vertebrates from GenBank Release 127.0 [15 December 2001]
\label{tabledata1}}
\footnotesize
\begin{center}
\begin{tabular}{|r|l|c|c|}
\hline
& Biological species & number of sequences & number of codons \\
\hline
1 & Homo sapiens & 41504 & 18611700 \\
2 & Mus musculus & 17286 & 8079821 \\
3 & Rattus norvegicus & 6578 & 3324518 \\
4 & Gallus gallus & 2089 & 1019029 \\
5 & Xenopus laevis & 1961 & 929562 \\
6 & Bos taurus & 1694 & 764195 \\
7 & Danio rerio & 1209 & 535583 \\
8 & Oryctolagus cuniculus & 864 & 441547 \\
9 & Macaca fascicularis & 1321 & 403875 \\
10 & Sus scrofa & 914 & 380357 \\
11 & Canis familiaris & 489 & 229526 \\
12 & Takifugu rubripes & 259 & 152479 \\
13 & Ovis aries & 413 & 134027 \\
14 & Oncorhynchus mykiss & 366 & 131431 \\
15 & Cricetulus griseus & 217 & 109395 \\
16 & Rattus sp. & 265 & 106164 \\
17 & Pan troglodytes & 272 & 88272 \\
18 & Oryzias latipes & 183 & 85610 \\
19 & Macaca mulatta & 264 & 81673 \\
20 & Felis cattus & 177 & 66930 \\
21 & Equus caballus & 177 & 59932 \\
\hline
\end{tabular}
\end{center}
\end{table}

\begin{table}[b]
\centering
\caption{Sum of usage probability of codons $P_{C+A}(XN) \equiv 
P(XNC)+P(XNA)$. The number in the first column denotes the biological 
species of Table \ref{tabledata1}. The amino acid are labelled by the 
standard letter. Morevover $P'_{C+A}(S) = P(UCA) + P(AGC)$.}
\label{tablePCA}
\footnotesize \medskip
\begin{tabular}{|c||c|c|c|c|c|c|c|c|c|}
\hline
\begin{tabular}{c} Biological \\ species \end{tabular}
& $P_{C+A}(P)$ & $P_{C+A}(A)$ & $P_{C+A}(T)$ & $P_{C+A}(S)$ & $P_{C+A}(V) $ 
& $P_{C+A}(L)$ & $P_{C+A}(R)$ & $P_{C+A}(G)$ & $P'_{C+A}(S)$ \\
\hline
 1 & 0.60 & 0.63 & 0.64 & 0.60 & 0.35 & 0.33 & 0.51 & 0.59 & 0.65 \\
 2 & 0.59 & 0.61 & 0.65 & 0.59 & 0.36 & 0.34 & 0.52 & 0.59 & 0.65 \\
 3 & 0.59 & 0.62 & 0.65 & 0.60 & 0.37 & 0.34 & 0.52 & 0.59 & 0.65 \\
 4 & 0.60 & 0.59 & 0.62 & 0.59 & 0.35 & 0.31 & 0.53 & 0.58 & 0.67 \\
 5 & 0.60 & 0.60 & 0.62 & 0.56 & 0.38 & 0.33 & 0.50 & 0.58 & 0.62 \\
 6 & 0.60 & 0.63 & 0.65 & 0.60 & 0.35 & 0.33 & 0.52 & 0.60 & 0.65 \\
 7 & 0.53 & 0.56 & 0.61 & 0.56 & 0.34 & 0.32 & 0.55 & 0.63 & 0.63 \\
 8 & 0.61 & 0.65 & 0.64 & 0.63 & 0.35 & 0.33 & 0.55 & 0.61 & 0.66 \\
 9 & 0.60 & 0.63 & 0.63 & 0.60 & 0.37 & 0.35 & 0.50 & 0.58 & 0.65 \\
10 & 0.61 & 0.64 & 0.65 & 0.62 & 0.36 & 0.33 & 0.53 & 0.61 & 0.67 \\
11 & 0.61 & 0.62 & 0.64 & 0.60 & 0.38 & 0.34 & 0.52 & 0.59 & 0.65 \\
12 & 0.58 & 0.58 & 0.61 & 0.59 & 0.38 & 0.32 & 0.55 & 0.59 & 0.64 \\
13 & 0.61 & 0.63 & 0.66 & 0.60 & 0.35 & 0.35 & 0.55 & 0.61 & 0.67 \\
14 & 0.62 & 0.61 & 0.68 & 0.60 & 0.38 & 0.33 & 0.53 & 0.57 & 0.66 \\
15 & 0.61 & 0.62 & 0.66 & 0.58 & 0.37 & 0.33 & 0.51 & 0.59 & 0.64 \\
16 & 0.58 & 0.62 & 0.66 & 0.59 & 0.37 & 0.34 & 0.52 & 0.59 & 0.66 \\
17 & 0.62 & 0.54 & 0.72 & 0.59 & 0.29 & 0.35 & 0.58 & 0.58 & 0.69 \\
18 & 0.56 & 0.59 & 0.63 & 0.60 & 0.35 & 0.31 & 0.55 & 0.63 & 0.68 \\
19 & 0.61 & 0.61 & 0.65 & 0.61 & 0.34 & 0.34 & 0.50 & 0.59 & 0.64 \\
20 & 0.61 & 0.63 & 0.64 & 0.60 & 0.38 & 0.34 & 0.52 & 0.60 & 0.64 \\
21 & 0.58 & 0.63 & 0.66 & 0.62 & 0.37 & 0.34 & 0.53 & 0.61 & 0.68 \\
\hline
\end{tabular}
\end{table}

\clearpage

\begin{table}[t]
\centering
\caption{Mean value, standard deviation and their ratio for the 
probabilities $P(XZN)$ corresponding to the eight amino-acids related to 
quartets or sextets for the choice of biological species of Table 
\ref{tabledata1}.}
\label{tablestatXZN}
\footnotesize \medskip
\begin{tabular}{|c||c|c|c|c|c|c|c|c|}
\hline
& $P(CCU)$ & $P(CCC)$ & $P(CCA)$ & $P(CCG)$ & $P(ACU)$ & $P(ACC)$ & 
$P(ACA)$ & $P(ACG)$ \\
\hline
$\overline{x}$ & 0.28 & 0.33 & 0.26 & 0.13 & 0.23 & 0.39 & 0.26 & 0.13 \\
$\sigma$ & 0.028 & 0.043 & 0.034 & 0.028 & 0.030 & 0.050 & 0.034 & 0.027 \\
$\sigma/\overline{x}$ & 10.0 \% & 12.8 \% & 13.3 \% & 22.3 \% & 13.1 \% & 
13.0 \% & 13.0 \% & 21.4 \% \\
\hline
\hline
& $P(GCU)$ & $P(GCC)$ & $P(GCA)$ & $P(GCG)$ & $P(UCU)$ & $P(UCC)$ & 
$P(UCA)$ & $P(UCG)$ \\
\hline
$\overline{x}$ & 0.27 & 0.40 & 0.21 & 0.12 & 0.30 & 0.38 & 0.22 & 0.10 \\
$\sigma$ & 0.026 & 0.046 & 0.035 & 0.029 & 0.027 & 0.036 & 0.026 & 0.020 \\
$\sigma/\overline{x}$ & 9.5 \% & 11.6 \% & 16.5 \% & 25.3 \% & 8.8 \% & 9.3 
\% & 12.0 \% & 20.2 \% \\
\hline
\hline
& $P(GUU)$ & $P(GUC)$ & $P(GUA)$ & $P(GUG)$ & $P(CUU)$ & $P(CUC)$ & 
$P(CUA)$ & $P(CUG)$ \\
\hline
$\overline{x}$ & 0.17 & 0.26 & 0.10 & 0.47 & 0.15 & 0.25 & 0.08 & 0.52 \\
$\sigma$ & 0.036 & 0.023 & 0.026 & 0.045 & 0.034 & 0.018 & 0.017 & 0.035 \\
$\sigma/\overline{x}$ & 20.9 \% & 9.1 \% & 25.8 \% & 9.5 \% & 22.6 \% & 7.2 
\% & 20.6 \% & 6.7 \% \\
\hline
\hline
& $P(CGU)$ & $P(CGC)$ & $P(CGA)$ & $P(CGG)$ & $P(GGU)$ & $P(GGC)$ & 
$P(GGA)$ & $P(GGG)$ \\
\hline
$\overline{x}$ & 0.16 & 0.34 & 0.18 & 0.31 & 0.17 & 0.33 & 0.26 & 0.23 \\
$\sigma$ & 0.042 & 0.039 & 0.026 & 0.043 & 0.029 & 0.034 & 0.033 & 0.032 \\
$\sigma/\overline{x}$ & 26.0 \% & 11.4 \% & 14.0 \% & 13.9 \% & 16.9 \% & 
10.3 \% & 12.7 \% & 13.7 \% \\
\hline
\end{tabular}
\end{table}

\begin{table}[t]
\centering
\caption{Mean value, standard deviation and their ratio for the sums of 
probabilities $P_{C+A}$, $P_{C+U}$, $P_{C+G}$ corresponding to the eight 
amino-acids related to quartets or sextets for the choice of biological 
species of Table \ref{tabledata1}. The amino acid are labelled by the 
standard letter.}
\label{tablestat}
\footnotesize \medskip
\begin{tabular}{|c||c|c|c|c|c|c|c|c|}
\hline
& $P_{C+A}(P)$ & $P_{C+A}(A)$ & $P_{C+A}(T)$ & $P_{C+A}(S)$ & $P_{C+A}(V) $ 
& $P_{C+A}(L)$ & $P_{C+A}(R)$ & $P_{C+A}(G)$ \\
\hline
$\overline{x}$ & 0.595 & 0.611 & 0.646 & 0.598 & 0.359 & 0.334 & 0.527 & 
0.596 \\
$\sigma$ & 0.020 & 0.027 & 0.024 & 0.016 & 0.020 & 0.012 & 0.020 & 0.015 \\
$\sigma/\overline{x}$ & 3.4 \% & 4.4 \% & 3.8 \% & 2.6 \% & 5.6 \% & 3.7 \% 
& 3.8 \% & 2.5 \% \\
\hline
\hline
& $P_{C+U}(P)$ & $P_{C+U}(A)$ & $P_{C+U}(T)$ & $P_{C+U}(S)$ & $P_{C+U}(V) $ 
& $P_{C+U}(L)$ & $P_{C+U}(R)$ & $P_{C+U}(G)$ \\
\hline
$\overline{x}$ & 0.613 & 0.672 & 0.614 & 0.687 & 0.430 & 0.401 & 0.506 & 
0.507 \\
$\sigma$ & 0.030 & 0.028 & 0.027 & 0.026 & 0.031 & 0.022 & 0.049 & 0.024 \\
$\sigma/\overline{x}$ & 4.9 \% & 4.2 \% & 4.4 \% & 3.8 \% & 7.2 \% & 5.4 \% 
& 9.7 \% & 4.8 \% \\
\hline
\hline
& $P_{C+G}(P)$ & $P_{C+G}(A)$ & $P_{C+G}(T)$ & $P_{C+G}(S)$ & $P_{C+G}(V) $ 
& $P_{C+G}(L)$ & $P_{C+G}(R)$ & $P_{C+G}(G)$ \\
\hline
$\overline{x}$ & 0.462 & 0.513 & 0.511 & 0.479 & 0.728 & 0.769 & 0.652 & 
0.567 \\
$\sigma$ & 0.058 & 0.058 & 0.063 & 0.046 & 0.056 & 0.048 & 0.055 & 0.058 \\
$\sigma/\overline{x}$ & 12.4 \% & 11.3 \% & 12.3 \% & 9.6 \% & 7.7 \% & 6.3 
\% & 8.4 \% & 10.2 \% \\
\hline
\end{tabular}
\end{table}

\end{document}